\documentclass[published]{JHEP3} 

\JHEP{00(2002)000}

\JHEPspecialurl{http://jhep.sissa.it/JOURNAL/JHEP3.tar.gz}

\usepackage{epsfig,multicol}

\newcommand\fverb{\setbox\pippobox=\hbox\bgroup\verb}
\newcommand\fverbdo{\egroup\medskip\noindent%
			\fbox{\unhbox\pippobox}\ }
\newcommand\fverbit{\egroup\item[\fbox{\unhbox\pippobox}]}
\newbox\pippobox

\def\sst{\scriptscriptstyle}

\def\a{\alpha}
\def\b{\beta}
\def\g{\gamma}
\def\d{\delta}
\def\D{\Delta}
\def\e{\epsilon}
\def\z{\zeta}
\def\t{\theta}
\def\T{\Theta}
\def\l{\lambda}
\def\o{\omega}

\def\be{\begin{equation}}
\def\ee{\end{equation}}
\def\ba{\begin{eqnarray}}
\def\ea{\end{eqnarray}}

\newcommand{\rpar}{\stackrel{\leftarrow}{\partial}}
\newcommand{\lpar}{\stackrel{\rightarrow}{\partial}}

\newcommand{\no}{\nonumber}

\title{Exterior Differentials in Superspace and Poisson Brackets}

\author{Dmitrij V. Soroka and Vyacheslav A. Soroka\\
	Kharkov Institute of Physics and Technology, 61108 Kharkov, Ukraine\\
	E-mail: \email{dsoroka@kipt.kharkov.ua}, 
        \email{vsoroka@kipt.kharkov.ua}}
\received{}
\accepted{}

\preprint{\hepth{0211280}}	

\abstract{It is shown that two definitions for an exterior
differential in superspace, giving the same exterior calculus, yet lead to 
different results when applied to the Poisson bracket. 
A prescription for the transition with the help of these exterior 
differentials from the given Poisson bracket of definite Grassmann 
parity to another bracket is introduced. 
It is also indicated that the resulting bracket 
leads to generalization of the Schouten-Nijenhuis bracket for the cases 
of superspace and brackets of  diverse Grassmann parities.
It is shown that in the case 
of the Grassmann-odd exterior differential  the resulting bracket is the 
bracket given on exterior forms. 
The above-mentioned transition with the use of the odd exterior differential
applied to the linear even/odd Poisson 
brackets, that correspond to semi-simple Lie groups, results, respectively, 
in also linear odd/even brackets which are naturally connected 
with the Lie superalgebra. The latter 
contains the BRST and anti-BRST charges and can be used for calculation 
of the BRST operator cohomology.}


\dedicated{Dedicated to the memory of Aleksandr Ivanovich Soroka and 
Raisa Artemovna Soroka (Gelukh)}

\begin{document}


\section{Introduction: two definitions of an exterior differential in
superspace}

There exist two possibilities to define an exterior
differential in superspace with coordinates $z^a$ having Grassmann
parities $g(z^a)\equiv g_a$ and satisfying the permutation relations
\ba
z^az^b=(-1)^{g_ag_b}z^bz^a.\no%
\ea
The first one realized when we set the Grassmann parity of the exterior
differential $d_{\sst0}=d_{\sst0}z^a\partial_{z^a}$ to be equal to zero
\ba
g(d_{\sst0}z^a)=g_a,\no
\ea
where $\partial_{z^a}\equiv\partial/\partial z^a$. In this case
the symmetry property of an exterior product of two differentials is
\ba\label{1}
d_{\sst 0}z^a\wedge d_{\sst 0}z^b=
(-1)^{g_ag_b+1}d_{\sst0}z^b\wedge d_{\sst0}z^a
\ea
and a permutation relation of the exterior differential $d_{\sst0}z^a$
with the coordinate $z^b$ has the form
\ba
d_{\sst0}z^az^b=(-1)^{g_ag_b}z^bd_{\sst0}z^a. \no
\ea
Note that relation (\ref{1}) can be rewritten in the following form
\ba\label{2}
(-1)^{g_a}d_{\sst 0}z^a\wedge d_{\sst 0}z^b=
(-1)^{(g_a+1)(g_b+1)}(-1)^{g_b}d_{\sst0}z^b\wedge d_{\sst0}z^a.
\ea
By defining an exterior product of a differential $p$-form
\ba
\Phi^0=d_{\sst 0}z^{a_p}\wedge\cdots\wedge d_{\sst 0}z^{a_1}
\phi^0_{{a_1}\ldots {a_p}},\qquad
g(\phi^0_{{a_1}\ldots {a_p}})=g_{a_1}+\cdots+g_{a_p}\no
\ea
and a $q$-form
\ba
\Psi^0=d_{\sst 0}z^{b_q}\wedge\cdots\wedge d_{\sst 0}z^{b_1}
\psi^0_{{b_1}\ldots {b_q}},\qquad
g(\psi^0_{{b_1}\ldots {b_q}})=g_{b_1}+\cdots+g_{b_q}\no
\ea
as
\ba
\Phi^0\wedge\Psi^0=(-1)^{pq}
d_{\sst 0}z^{b_q}\wedge\cdots\wedge d_{\sst 0}z^{b_1}\wedge
d_{\sst 0}z^{a_p}\wedge\cdots\wedge d_{\sst 0}z^{a_1}
\phi^0_{{a_1}\ldots {a_p}}\psi^0_{{b_1}\ldots {b_q}},\no
\ea
we obtain the following symmetry property of this product
\ba
\Phi^0\wedge\Psi^0=(-1)^{pq}\Psi^0\wedge\Phi^0.\no
\ea
By setting the exterior differential of the $p$-form $\Phi^0$ as follows
\ba\label{3}
d_{\sst 0}\Phi^0=
d_{\sst 0}\wedge d_{\sst 0}z^{a_p}\wedge\cdots\wedge d_{\sst 0}z^{a_1}
\phi^0_{{a_1}\ldots {a_p}}=
(-1)^pd_{\sst 0}z^{a_p}\wedge\cdots\wedge d_{\sst 0}z^{a_1}\wedge
d_{\sst0}z^b\partial_{z^b}\phi^0_{{a_1}\ldots {a_p}},
\ea
we obtain the Leibnitz rule for the exterior differential of the exterior
product of two forms
\ba
d_{\sst 0}(\Phi^0\wedge\Psi^0)=(d_{\sst 0}\Phi^0)\wedge\Psi^0+
(-1)^p\Phi^0\wedge(d_{\sst 0}\Psi^0).\no
\ea
Note that very often another definition for the exterior differential is
adopted
\ba
\overline{d_{\sst 0}\Phi^0}=
d_{\sst 0}z^{a_1}\wedge\cdots\wedge d_{\sst 0}z^{a_p}\wedge
d_{\sst0}z^b\partial_{z^b}\phi^0_{{a_p}\ldots {a_1}}\no
\ea
which differs from (\ref{3}) with the absence of the grading factor
$(-1)^p$ and leads to the following form of the Leibnitz rule
\ba
\overline{d_{\sst 0}(\Phi^0\wedge\Psi^0)}=
(-1)^q(\overline{d_{\sst 0}\Phi^0})\wedge\Psi^0+
\Phi^0\wedge(\overline{d_{\sst 0}\Psi^0}).\no
\ea

Another definition for the exterior differential in superspace arises when
the Grassmann parity of the exterior differential $d_{\sst1}=
d_{\sst1}z^a\partial_{z^a}$ is chosen to be equal to unit
\ba
g(d_{\sst1}z^a)=g_a+1.\no
\ea
Then the symmetry property of the exterior product for two differentials is
defined as\footnote{In this case we use another notation $\tilde\wedge$
for the exterior multiplication.}
\ba\label{4}
d_{\sst1}z^a\tilde\wedge d_{\sst1}z^b=
(-1)^{(g_a+1)(g_b+1)}d_{\sst1}z^b\tilde\wedge d_{\sst1}z^a
\ea
and a rule for the permutation of such a differential with the coordinate
$z^b$ has to be
\ba
d_{\sst1}z^az^b=(-1)^{(g_a+1)g_b}z^bd_{\sst1}z^a. \no
\ea
Relation (\ref{4}) can be represented in the form
\ba\label{5}
(-1)^{g_a}d_{\sst 1}z^a\tilde\wedge d_{\sst 1}z^b=
(-1)^{g_ag_b+1}(-1)^{g_b}d_{\sst1}z^b\tilde\wedge d_{\sst1}z^a.
\ea
If the exterior product of the $p$-form
\ba
\Phi^1=d_{\sst 1}z^{a_p}\tilde\wedge\cdots\tilde\wedge d_{\sst 1}z^{a_1}
\phi^1_{{a_1}\ldots {a_p}},\qquad
g(\phi^1_{{a_1}\ldots {a_p}})=g_{a_1}+\cdots+g_{a_p}\no
\ea
and $q$-form
\ba
\Psi^1=d_{\sst 1}z^{b_q}\tilde\wedge\cdots\tilde\wedge d_{\sst 1}z^{b_1}
\psi^1_{{b_1}\ldots {b_q}},\qquad
g(\psi^1_{{b_1}\ldots {b_q}})=g_{b_1}+\cdots+g_{b_q}\no
\ea
is defined in the following way
\ba
\Phi^1\tilde\wedge\Psi^1=(-1)^{p(q+g_{b_1}+\cdots+g_{b_q})}
d_{\sst 1}z^{b_q}\tilde\wedge\cdots\tilde\wedge d_{\sst 1}z^{b_1}\tilde\wedge
d_{\sst 1}z^{a_p}\tilde\wedge\cdots\tilde\wedge d_{\sst 1}z^{a_1}
\phi^1_{{a_1}\ldots {a_p}}\psi^1_{{b_1}\ldots {b_q}},\no
\ea
then the symmetry property of this product is
\ba
\Phi^1\tilde\wedge\Psi^1=(-1)^{pq}\Psi^1\tilde\wedge\Phi^1.\no
\ea
Let us define in this case the exterior differential of the $p$-form
$\Phi^1$ in the following form
\ba
d_{\sst 1}\Phi^1=
d_{\sst 1}\tilde\wedge d_{\sst 1}z^{a_p}\tilde\wedge\cdots
\tilde\wedge d_{\sst 1}z^{a_1}\phi^1_{{a_1}\ldots {a_p}}=
(-1)^{p+g_{a_1}+\cdots+g_{a_p}}
d_{\sst 1}z^{a_p}\tilde\wedge\cdots\tilde\wedge d_{\sst 1}z^{a_1}\tilde\wedge
d_{\sst1}z^b\partial_{z^b}\phi^1_{{a_1}\ldots {a_p}}.\no
\ea
Then the Leibnitz rule for the exterior differential of the exterior
product of a $p$-form $\Phi^1$ and a $q$-form $\Psi^1$ will be
\ba
d_{\sst 1}(\Phi^1\tilde\wedge\Psi^1)=(d_{\sst 1}\Phi^1)\tilde\wedge\Psi^1+
(-1)^p\Phi^1\tilde\wedge(d_{\sst 1}\Psi^1).\no
\ea
Note that in this case the symmetry properties of the exterior product
(\ref{4}) coincide with the ones for the usual Grassmann product of two
differentials for the coordinate $z^a$
\ba\label{6}
d_{\sst1}z^ad_{\sst1}z^b=
(-1)^{(g_a+1)(g_b+1)}d_{\sst1}z^bd_{\sst1}z^a.
\ea

The equivalence of the exterior calculi obtained with the use of the above
mentioned different definitions for the exterior differential can be
established as a result of the direct verification by taking into account
relations (\ref{1}) and (\ref{5}) and by putting the following relations
between coefficients of the corresponding $p$-forms $\Phi^0$ and $\Phi^1$
\ba
\phi^0_{{a_1}\ldots{a_p}}=(-1)^{\sum\limits_{k=1}^{[p/2]}
g_{a_{\sst2\sst k}}} \phi^1_{{a_1}\ldots{a_p}},\no
\ea
where $[p/2]$ denotes a whole part of the quantity $p/2$.
Thus, we proved that two definitions for the exterior differential,
differed with the Grassmann parities, result in the same exterior calculus.

\section{Poisson brackets related with the exterior differentials}

Now we show that application of these differentials leads, however, to the
different results under construction from a given Poisson bracket with a
Grassmann parity $\e=0,1$ of another one.

A Poisson bracket, having a Grassmann parity $\e$, written in arbitrary
non-canonical variables $z^a$
\ba\label{in}
\{A,B\}_\e=A\rpar_{z^a}\o_\e^{ab}(z)\lpar_{z^b}B
\ea
has the following main properties:
\ba
g(\{A,B\}_\e)\equiv g_A+g_B+\e\pmod2,\no
\ea
\ba
\{A,B\}_\e=-(-1)^{(g_A+\e)(g_B+\e)} \{B,A\}_\e,\no
\ea
\ba
\sum_{(ABC)}(-1)^{(g_A+\e)(g_C+\e)} \{A,\{B,C\}_\e\}_\e=0,\no
\ea
which lead to the corresponding relations for the matrix
$\o_\e^{ab}$
\be
g\left(\o_\e^{ab}\right)\equiv g_a+g_b+\e\pmod2,\label{8}
\ee
\be
\o_\e^{ab}=-(-1)^{(g_a+\e)(g_b+\e)} \o_\e^{ba},\label{9}
\ee
\be
\sum_{(abc)}(-1)^{(g_a+\e)(g_c+\e)} \o_\e^{ad}\partial_{z^d}
\o_\e^{bc}=0,\label{10}
\ee
where $g_A\equiv g(A)$ and a sum with a symbol $(abc)$ under it
designates a summation over cyclic permutations of $a, b$ and $c$.

The Hamilton equations for the phase variables $z^a$, which correspond
to a Hamiltonian $H_\e$ ($g(H_\e)=\e$),
\ba
{dz^a\over dt}=\{z^a,H_\e\}_\e=
\o_\e^{ab}\partial_{z^b}H_\e\label{11}
\ea
can be represented in the form
\ba
{dz^a\over dt}=\o_\e^{ab}\partial_{z^b}H_\e\equiv
\o_\e^{ab}{\partial(d_{\sst\z}H_\e)\over\partial(d_{\sst\z}z^b)}
\mathrel{\mathop=^{\rm def}}(z^a,d_{\sst\z}H_\e)_{\e+\z},\no
\ea
where $d_{\sst\z}$ ($\z=0,1$) is one of the exterior differentials
$d_{\sst0}$ or $d_{\sst1}$. By taking the exterior differential
$d_{\sst\z}$ from the Hamilton equations (\ref{11}), we obtain
\ba
{d(d_{\sst\z}z^a)\over dt}=(d_{\sst\z}\o_\e^{ab})
{\partial(d_{\sst\z}H_\e)\over\partial(d_{\sst\z}z^b)}
+(-1)^{\z(g_a+\e)}\o_\e^{ab}\partial_{z^b}(d_{\sst\z}H_\e)
\mathrel{\mathop=^{\rm def}}(d_{\sst\z}z^a,d_{\sst\z}H_\e)_{\e+\z}.\no
\ea
As a result of two last equations we have by definition the following
binary composition for functions $F$ and $G$ of the variables $z^a$ and
their differentials $d_{\sst\z}z^a\equiv y_{\sst\z}^a$
\ba\label{fin}
(F,G)_{\e+\z}=
F\left[\rpar_{z^a}\o_\e^{ab}\lpar_{y_{\sst\z}^b}+
(-1)^{\z(g_a+\e)}\rpar_{y_{\sst\z}^a}
\o_\e^{ab}\lpar_{z^b}+
\rpar_{y_{\sst\z}^a}y_{\sst\z}^c\left(\partial_{z^c}
\o_\e^{ab}\right)\lpar_{y_{\sst\z}^b}\right]G.
\ea
In consequence of the grading properties (\ref{8}) for the matrix
$\o_\e^{ab}$ this composition has the Grassmann parity $\e+\z$
\ba
g[(F,G)_{\e+\z}]\equiv g_F+g_G+\e+\z\pmod2.\no
\ea
By using the symmetry property (\ref{9}) of $\o_\e^{ab}$ , we can
establish the symmetry of the composition (\ref{fin})
\ba
(F,G)_{\e+\z}=-(-1)^{(g_F+\e+\z)(g_G+\e+\z)} (G,F)_{\e+\z}.\no
\ea
At last, taking into account relations (\ref{9}) and (\ref{10}) for the
matrix $\o_\e^{ab}$, we come to the Jacobi identities for this composition
\ba
\sum_{(EFG)}(-1)^{(g_E+\e+\z)(g_G+\e+\z)} (E,(F,G)_{\e+
\z})_{\e+\z}=0.\no
\ea

We see that the composition (\ref{fin}) satisfies all the main properties
for the Poisson bracket with the Grassmann parity equal to $\e+\z$. Thus,
the application of the exterior differentials of opposite Grassamann
parities to the given Poisson bracket results in the brackets of the
different Grassmann parities.

Note that by transition to the variables $y^{\sst\e+\sst\z}_a$, related
with $y_{\sst\z}^a$ by means of the matrix $\o_\e^{ab}$
\ba\label{13}
y_{\sst\z}^a=y^{\sst\e+\sst\z}_b\o_\e^{ba},
\ea
the Poisson bracket (\ref{fin}) takes a canonical form\footnote{There is 
no summation over $\e$ in relation (\ref{13}).}
\ba
(F,G)_{\e+\z}=F\left[\rpar_{z^a}\lpar_{y^{\sst\e+\sst\z}_a}
-(-1)^{g_a(g_a+\e+\z)}\rpar_{y^{\sst\e+\sst\z}_a}\lpar_{z^a}\right]G\no
\ea
that can be proved with the use of the Jacobi identities (\ref{10}).

In the case $\z=1$, due to relations (\ref{4}), (\ref{6}), the terms in the
decomposition of a function $F(z^a,y_{\sst1}^a)$ into degrees $p$ of the
variables $y_{\sst1}^a$ can be treated as $p$-forms and the bracket
(\ref{fin}) can be considered as a Poisson bracket on $p$-forms so that
being taken between a $p$-form and a $q$-form results in a
$(p+q-1)$-form\footnote{Concerning Poisson bracket between 1-forms and its 
relation with Lie bracket of vector fields see in the book~\cite{stern}.}. 
The bracket (\ref{fin}) is a generalization of the
bracket introduced in~\cite{kar,karmas} on the superspace case and on the 
case of the brackets (\ref{in}) with arbitrary Grassmann parities.

Let us also note that if we take the bracket (\ref{fin}) in the component 
form and rise low indexes with the use of the matrix $\o_\e^{ab}$ according 
to the rule (\ref{13}) then we come to the generalizations of the 
Schouten-Nijenhuis brackets~\cite{scho,nij} 
(see also~\cite{karmas,nij1,fr-nij,kod-sp,but,bffls,oz}) 
onto the cases of superspace and the brackets 
of diverse Grassmann parities. The details of this generalization will be 
given in a separate paper \cite{prep}.

It follows from the structure of the bracket (\ref{fin}) that if the
initial bracket (\ref{in}) is degenerate and possessed of a Casimir
function $C(z)$
\ba
\{\ldots,C\}_{\e}=0,\no
\ea
then the bracket (\ref{fin}) has as Casimir functions this one
\ba
(\ldots,C)_{\e+\z}=0\no
\ea
and also a function of the form
\ba\label{15}
\tilde C=y_{\sst\z}^a\partial_{z^a}C,\quad(\ldots,\tilde C)_{\e+\z}=0.
\ea

\section{Linear Poisson brackets related with semi-simple Lie groups}

Here we apply the procedure described in the previous section to the
linear even and odd brackets connected with a semi-simple Lie group having
structure constants ${c_{\a\b}}^\g$ which obey the usual conditions
\ba
{c_{\a\b}}^\g=-{c_{\b\a}}^\g,\no
\ea
\ba
\sum_{(\a\b\g)}{c_{\a\b}}^\l{c_{\l\g}}^\d=0.\no
\ea

Let us take as an initial Poisson bracket (\ref{in})
the linear even bracket given in terms of the commuting variables 
\footnote{Lie-Poisson-Kirillov bracket.} $x_\a$ (here $z^a=x_\a$)
\ba
\{A,B\}_0=A\rpar_{x_\a}{c_{\a\b}}^\g
x_\g\lpar_{x_\b}B.\label{even}
\ea
In the case of a semi-simple Lie group, which hereafter will be
considered, this bracket has a Casimir function
\ba
C_0=x_\a x_\b g^{\a\b},\quad\{\ldots,C_0\}_0=0,\label{C0}
\ea
where $g^{\a\b}$ is an inverse tensor to the Cartan--Killing metric
\ba
g_{\a\b}={c_{\a\g}}^\l{c_{\b\l}}^\g.\no
\ea
By using the odd exterior differential $d_{\sst1}$, we obtain from the
bracket (\ref{even}) in conformity with the transition from the bracket
(\ref{in}) to the bracket (\ref{fin}) the following linear odd bracket
\ba
(F,G)_1=F(\rpar_{x_\a}{c_{\a\b}}^\g
x_\g\lpar_{\t_\b}+
\rpar_{\t_\a}{c_{\a\b}}^\g
x_\g\lpar_{x_\b}+
\rpar_{\t_\a}{c_{\a\b}}^\g
\t_\g\lpar_{\t_\b})G,\label{odd}
\ea
where $\t_\a=d_{\sst1}x_\a$ are Grassmann variables. In this case relation
(\ref{13}) has the form
\ba
\t_\a=\T^\b{c_{\b\a}}^\g x_\g,\no
\ea
where $\T^\b$ are also Grassmann variables in term of which the odd bracket
(\ref{odd}) takes a canonical form
\ba
(F,G)_1=F(\rpar_{x_\a}\lpar_{\T^\a}-\rpar_{\T^\a}\lpar_{x_\a})G.\no
\ea
According to (\ref{15}) the bracket (\ref{odd}) has as Casimir
functions apart from $C_0$ (\ref{C0}) a nilpotent quantity
\ba
\tilde C_1=x_\a\t_\b g^{\a\b},\quad(\ldots,\tilde C_1)_1=0,
\quad(\tilde C_1)^2=0.\no
\ea

The odd bracket (\ref{odd}) has two nilpotent Batalin-Vilkovisky type 
differential second order $\D$-operators~\cite{bv1,bv2} 
(see also~\cite{schw,schw1,kn})
\ba
\D_{-1}=-{1\over2}[\partial_{x_\a}(x_\a, )_1+
\partial_{\t_\a}(\t_\a, )_1]=-{1\over2}S_\a\partial_{\t_\a},
\quad (\D_{-1})^2=0\label{-1}
\ea
and
\ba
\D=-{1\over2}[\partial_{x_\a}(x_\a, )_1-
\partial_{\t_\a}(\t_\a, )_1]=
(T_\a+{1\over2}S_\a)\partial_{\t_\a},\quad
\D^2=0,\label{D}
\ea
where
\ba
T_\a={c_{\a\b}}^\g x_\g\partial_{x_\b}\label{T}
\ea
and
\ba
S_\a={c_{\a\b}}^\g\t_\g\partial_{\t_\b}\label{S}
\ea
are generators of the Lie group in the co-adjoint representation which
obey the commutation relations
\ba
[T_\a,T_\b]={c_{\a\b}}^\g T_\g,\no
\ea
\ba
[S_\a,S_\b]={c_{\a\b}}^\g S_\g,\no
\ea
\ba
[T_\a,S_\b]=0.\no
\ea
Note that
\ba
(\t_\a, )_1=T_\a+S_\a\equiv Z_\a\no
\ea
and $Z_\a$ satisfy the relations
\ba
[Z_\a,Z_\b]={c_{\a\b}}^\g Z_\g.\label{Z}
\ea

Now let us take as an initial bracket (\ref{in}) the linear odd bracket
introduced in Refs.~\cite{s,ss} and given in terms of Grassmann variables
$\t_\a$ (in this case $z^a=\t_\a$)
\ba
\{A,B\}_1=A\rpar_{\t_\a}{c_{\a\b}}^\g
\t_\g\lpar_{\t_\b}B,\label{odd1}
\ea
which has in the case of the semi-simple Lie group a nilpotent Casimir
function
\ba
C_1=\t^\a\t^\b\t^\g c_{\a\b\g},\quad\{\ldots,C_1\}_1=0,
\quad (C_1)^2=0,\label{C1}
\ea
where $\t^\a=g^{\a\b}\t_\b$ and $c_{\a\b\g}={c_{\a\b}}^\l g_{\l\g}$.
With the help of the odd differential $d_{\sst1}$, according to the
transition from (\ref{in}) to (\ref{fin}), we come from the bracket
(\ref{odd1}) to the even linear bracket of the form
\ba
(F,G)_0=F(\rpar_{\t_\a}{c_{\a\b}}^\g
\t_\g\lpar_{x_\b}+
\rpar_{x_\a}{c_{\a\b}}^\g
\t_\g\lpar_{\t_\b}+
\rpar_{x_\a}{c_{\a\b}}^\g
x_\g\lpar_{x_\b})G,\label{even1}
\ea
where $x_\a=d_{\sst1}\t_\a$ are commuting variables. Relation (\ref{13})
in this case has the form
\ba
x_\a=\T^\b{c_{\b\a}}^\g\t_\g,\no
\ea
where $\T^\b$ are Grassmann variables in term of which the bracket
(\ref{even1}) takes a canonical form for the Martin bracket \cite{mar}
\ba
(F,G)_0=F(\rpar_{\t_\a}\lpar_{\T^\a}+\rpar_{\T^\a}\lpar_{\t_\a})G.\no
\ea
In accordance with (\ref{15}) the even bracket (\ref{even1}) has as
Casimir functions besides $C_1$ (\ref{C1}) the function
\ba
\tilde C_0=x^\a\t^\b\t^\g c_{\a\b\g},\quad (\ldots,\tilde C_0)_0=0,\no
\ea
where $x^\a=g^{\a\b}x_\b$.

The even bracket (\ref{even1}), in contrast to the odd bracket, has no 
second order nilpotent differential $\D$-like operators. It is surprising 
enough that instead of this it has two {\it nilpotent}
differential operators of the first order
\ba
\D_1=-{1\over2}[\t^\a(x_\a, )_0+
x^\a(\t_\a, )_0]=
-{1\over2}\t^\a S_\a,\quad (\D_1)^2=0.
\ea
and
\ba
Q={1\over2}[\t^\a(x_\a, )_0-
x^\a(\t_\a, )_0]=
\t^\a(T_\a+{1\over2}S_\a),\quad Q^2=0.\label{Q}
\ea
In the papers \cite{s,ss} the operators $\D_1$ and $\D_{-1}$, defined 
on the Grassmann algebra with generators $\t_\a$, have been
introduced in connection with the linear odd Poisson bracket
(\ref{odd1}), which corresponds to a semi-simple Lie group, and the
Lie superalgebra for them has been given.  These operators are standard
terms in the BRST and anti-BRST charges respectively.

\section{Lie superalgebra for the BRST and anti-BRST charges}

Thus, in the superspace with coordinates $x_\a, \t_\a$ with the help of
the linear even (\ref{even1}) and odd (\ref{odd}) Poisson brackets we
constructed the operators $Q$ (\ref{Q}) and $\D$ (\ref{D}). These
operators can be treated as the BRST and anti-BRST charges correspondingly
(see, e.g.,~\cite{h}) if we consider $\t^\a$ and $\partial_{\t^\a}$ as
representations for the ghosts and antighosts operators respectively. The
operators $Q$ and $\D$ satisfy the following anticommutation relation:
\ba
\{Q,\D\}={1\over2}(T^\a T_\a+Z^\a Z_\a),\label{Q,D}
\ea
two terms in the right-hand side of which, because of the commutation
relations
\ba
[T^\a T_\a,Q]=0,\label{TT,Q}
\ea
\ba
[T^\a T_\a,\D]=0,\label{TT,D}
\ea
\ba
[Z_\a,Q]=0,\label{Z,Q}
\ea
\ba
[Z_\a,\D]=0,\label{Z,D}
\ea
\ba
[T^\a T_\a,Z_\b]=0,\label{TT,Z}
\ea
are central elements of the Lie superalgebra formed by the quantities $Q$,
$\D$, $T^\a T_\a$ and $Z^\a Z_\a$. The relations (\ref{Q,D})--(\ref{TT,Z})
remain valid if we take instead of the co-adjoint representation
(\ref{T}) an arbitrary representation for the generators $T_\a$. The
quantity $Z^\a Z_\a$ contains the term $S^\a S_\a$ which can be written as
\ba
S^\a S_\a=N-K\no
\ea
where
\ba
N=\t^\a\partial_{\t^\a},
\ea
can be considered as a representation for the ghost number operator and
the quantity $K$ has the form
\ba
K = {1\over 2} \t^\a \t^\b
{c_{\a\b}}^\l c_{\l\g\d}
\partial_{\t_\g} \partial_{\t_\d}.\no
\ea
The operator $N$ has the following permutation relations with $Q$ and $\D$:
\ba
[N,Q]=Q,\label{N,Q}
\ea
\ba
[N,\D]=-\D\label{N,D}
\ea
and commutes with the central elements $T^\a T_\a$ and $Z^\a Z_\a$
\ba
[N,T_\a]=0,\label{N,T}
\ea
\ba
[N,Z_\a]=0.\label{N,Z}
\ea
We can add the commutation relations (\ref{Z}) for the generators $Z_\a$
with the usual quadratic Casimir operator $Z^\a Z_\a$ for the semi-simple
Lie group.

Note that the Lie superalgebra for the quantities $Q$, $\D$, $N$,
$T^\a T_\a$ and $Z^\a Z_\a$ determined by the relations
(\ref{Z}), (\ref{Q,D})--(\ref{N,Z}) can be used for the calculation of the
BRST operator cohomologies~\cite{hv}.

\section{Conclusion}

Thus, we illustrated that in superspace the exterior differentials with
opposite Grassmann parities give the same exterior calculus.

Then we introduced a prescription for the construction with the help of
these differentials from a given Poisson bracket of the definite Grassmann
parity of another one. We showed that the parity of the resulting bracket
depends on the parities of the both initial bracket and exterior differential 
used.

It is also indicated that the resulting bracket is related with a
generalization of the Schouten-Nijenhuis bracket on the superspace case and 
on the brackets of an arbitrary Grassmann parity.

By applying the prescription to the linear odd and even Poisson brackets,
corresponding to a semi-simple Lie group with the structure constants
${c_{\a\b}}^\g$ and given respectively on the anticommuting $\t_\a$ and
commuting $x_\a$ variables, we come with the help of the
Grassmann-odd exterior differential to the correspondingly even and odd linear
Poisson brackets which are both defined on the superspace with the
coordinates $x_\a, \t_\a$ and related also with the same semi-simple Lie
group.

We demonstrated that these resulting even and odd brackets are naturally
connected with the BRST and anti-BRST charges respectively.

\section*{Acknowledgments}
One of the authors (V.A.S.) is grateful to  A.P. Isaev, J. Lukierski,
M. Tonin and J. Wess for useful discussions and to S.J. Gates, Jr., P. Van
Nieuwenhuizen, W. Siegel and B. Zumino for stimulating discussions and
for hospitality respectively at the University of Maryland, SUNY (Stony
Brook) and LBNL (Berkeley) where the parts of the work have been performed.
V.A.S. thanks L. Bonora for fruitful discussions and for hospitality at
SISSA (Trieste) where this work has been completed.

\end{document}